\documentclass[sigconf,nonacm]{acmart}

\AtBeginDocument{%
  }

\setcopyright{none}
\copyrightyear{2026}
\acmYear{2026}
\acmDOI{}
\acmConference[IoT 2026]{The 16th International Conference on the Internet of Things (IoT 2026)}{November 17--20, 2026}{Newcastle upon Tyne, UK}
\acmISBN{}
\usepackage{pgfplots}
\usepackage{pgfplotstable}
\usepackage{tikz}
\usepackage{afterpage}
\usepackage{xcolor}
\usepackage{graphicx}
\usepackage{svg}
\usepackage{siunitx}
\usepackage{listings}
\usepackage{subcaption}
\usepackage[printonlyused,nolist]{acronym}
\usepackage{algorithm}
\usepackage[noend]{algpseudocode}
\usepackage{amsmath}
\usepackage{xspace}

\definecolor{CustomPrimary}{HTML}{125A87}       
\definecolor{CustomAxes}{HTML}{263746}          
\definecolor{CustomAlert}{HTML}{D65353}         
\definecolor{CustomChart1}{HTML}{218C9B}        
\definecolor{CustomChart2}{HTML}{6FA8CF}        
\definecolor{CustomChart3}{HTML}{4F63B5}        

\newcounter{observation}
\newcommand{\observation}[2][]{%
  \refstepcounter{observation}%
  \par\addvspace{\medskipamount}\noindent
  \textbf{Observation \theobservation%
  \if\relax\detokenize{#1}\relax\else\ (#1)\fi:} \textit{#2}
  \par\addvspace{\medskipamount}%
}

\newcounter{finding}
\newcommand{\finding}[2][]{%
  \refstepcounter{finding}%
  \par\addvspace{\medskipamount}\noindent
  \textbf{Finding \thefinding%
  \if\relax\detokenize{#1}\relax\else\ (#1)\fi:} \textit{#2}
  \par\addvspace{\medskipamount}%
}

\begin{document}

\begin{acronym}
    \acro{JIT}{Just-in-Time}
    \acro{AoT}{Ahead-of-Time}
    \acro{LEO}{Low Earth Orbit}
    \acro{CRIU}{Checkpoint and Restore in Userspace}
    \acro{SLO}{Service-Level Objective}
    \acro{TVG}{Time-Varying Graph}
    \acro{WASM}{WebAssembly}
    \acro{WASI}{WebAssembly System Interface}
    \acro{WAMR}{WebAssembly Micro Runtime}
    \acro{XR}{Extended Reality}
    \acro{RTT}{Round-Trip-Time}
    \acro{ISL}{Inter-Satellite Link}
    \acro{GSL}{Ground-to-Satellite Link}
    \acro{COTS}{Commercial Off-The-Shelf}
    \acro{SEU}{Single-Event Upset}
    \acro{SBC}{Single-Board Computer}
    \acro{OEC}{Orbital Edge Computing}
    \acro{ISA}{Instruction Set Architecture}
    \acro{SGP4}{Simplified General Perturbations 4}
    \acro{VS}{Virtual Stationarity}
    \acro{VSB}{Virtual Stationarity Boundary}
    \acro{VSR}{Virtual Stationarity Runtime}
    \acro{MEC}{Mobile Edge Computing}
    \acro{M2P}{Motion-to-Photon}
\end{acronym}

\usetikzlibrary{patterns, positioning, fit, backgrounds, calc, arrows.meta, shapes.geometric}
\pgfplotsset{compat=1.18}

\usepgfplotslibrary{fillbetween}
\usepgfplotslibrary{groupplots}

\title[Migration as a Steady-State Execution Primitive: A Runtime for Virtual Stationarity in Low-Earth Orbit]{Migration as a Steady-State Execution Primitive:\\A Runtime for Virtual Stationarity in Low-Earth Orbit}

\author{Daniel Kleebinder}
\email{d.kleebinder@dsg.tuwien.ac.at}
\orcid{0009-0008-7251-0313}
\correspondingauthor
\affiliation{%
  \institution{Distributed Systems Group, TU Wien}
  \city{Vienna}
  \country{Austria}
}

\author{Alireza Furutanpey}
\email{a.furutanpey@dsg.tuwien.ac.at}
\orcid{0000-0001-5621-7899}
\affiliation{%
  \institution{Distributed Systems Group, TU Wien}
  \city{Vienna}
  \country{Austria}
}

\author{Stefan Nastic}
\email{s.nastic@dsg.tuwien.ac.at}
\orcid{0000-0003-0410-6315}
\affiliation{%
  \institution{Distributed Systems Group, TU Wien}
  \city{Vienna}
  \country{Austria}
}

\renewcommand{\shortauthors}{Kleebinder et al.}

\begin{abstract}
\ac{LEO} satellite mega-constellations enable orbital edge computing for latency-critical geocentric applications, but satellite motion compresses line-of-sight communication windows to mere minutes. To maintain logical geographic proximity -- a paradigm termed \acl{VS} -- applications must continuously migrate across passing satellites in the constellation. Traditional live migration mechanisms optimize individually initiated relocations or replication, but not the continuous handover between successive hosts that orbital motion requires. We reformulate live migration as a steady-state execution model and introduce a novel virtual stationarity runtime paradigm. We formalize the viability of continuous migration through a lightweight model capturing latency and availability \acsp{SLO}, continuous migration overhead, and the convergence bound of state replay. To satisfy these constraints, we derive a \ac{VSR} framework whose Migration Membrane combines Capability Virtualization and Continuous State Projection to rebind host resources and reconstruct evolving execution on a successor. Evaluation on physical heterogeneous hosts shows $p_{95}$ downtime below 6 ms across 300 recurrent handovers, migration time effectively invariant to memory-write rate, and no cumulative growth in migration time or continuation state; a 1,600-satellite simulation further confirms completion within orbital handover windows.
\end{abstract}

\ccsdesc[500]{Computer systems organization~Heterogeneous (hybrid) systems}
\ccsdesc[500]{Computer systems organization~Distributed architectures}
\ccsdesc[500]{Computer systems organization~Reliability}
\ccsdesc[300]{Networks~Network dynamics}

\keywords{edge computing, satellite networks, live migration}

\maketitle

\section{Introduction}
\label{sec:introduction}

By mid-2026, more than 16,200 active payloads were in orbit, more than twice the number recorded three years earlier~\cite{mcdowell_jonathans_2026}. This rapid proliferation is catalyzing a fundamental shift in spaceborne computing~\cite{denby_orbital_2020}: recent research increasingly envisions satellites not only as passive communication relays but as active compute nodes interconnected via optical \acp{ISL} and linked to the ground through \acp{GSL}, forming a unified distributed computing continuum that bridges space and Earth~\cite{bhattacherjee_-orbit_2020,denby_orbital_2020}. At altitudes between 80~km and 2{,}000~km~\cite{catherine_g_manning_scan_2023}, \ac{LEO} satellites yield single-digit-millisecond propagation delays to ground users~\cite{bhattacherjee_-orbit_2020}. Hosting latency-critical workloads onboard passing spacecraft can therefore exploit this physical proximity to satisfy stringent real-time \acsp{SLO}~\cite{meng_enabling_2023,rong_live_2024,pfandzelter_komet_2024}.

However, extreme satellite velocities ($\approx 27{,}000~\text{km/h}$) compress direct line-of-sight contact windows over target ground regions to mere minutes~\cite{zhao_demystifying_2026,bhattacherjee_network_2019}. To maintain logical proximity, applications must therefore continuously migrate across passing spacecraft at high frequencies, elevating live migration from an exceptional terrestrial operation~\cite{attar-khorasani_seamless_2025,he_taxonomy_2023} to a steady-state \ac{LEO} execution primitive. As §\ref{sec:motivation} demonstrates, canonical migration strategies fail during high-frequency handovers by causing prolonged downtime, non-convergence under write pressure, or lingering predecessor dependencies. Advanced terrestrial mechanisms rely on assumptions -- such as kernel homogeneity, deterministic behavior, or extended predecessor reachability~\cite{scheuer_resync_2026,nadgowda_voyager_2017,bergan_deterministic_2010} -- that collapse in orbit, where nodes differ in \ac{ISA} and payload capabilities, migration budgets are tightly bounded, and source connectivity terminates abruptly.

To align orbital transience with terrestrial stability, prior work introduced \ac{VS} to logically anchor applications over fixed ground targets~\cite{bhattacherjee_-orbit_2020}. We extend \ac{VS} from a placement objective into a runtime model that establishes live migration as a steady-state execution primitive. Specifically, we make three core contributions: (i)~we formalize the system model and feasibility conditions governing recurrent orbital migration, establishing bounds of latency and availability \acsp{SLO}, hosting windows, and cumulative steady-state overhead (§\ref{sec:methodology:formalization}); (ii)~we design a \ac{VSR} framework that separates external orchestration from runtime handover and defines a four-phase Prepare--Reconstruct--Project--Commit protocol for recurrent orbital migration (§\ref{sec:methodology:framework}); and (iii)~we introduce the \textit{Migration Membrane}, comprising two enabling mechanisms: Capability Virtualization ($\mathcal{V}$) to dynamically rebind host-dependent resources across heterogeneous satellite configurations, and Continuous State Projection ($\mathcal{P}$) to asynchronously prepare successor nodes while the source continues executing (§\ref{sec:methodology:membrane}). We instantiate the framework and membrane within \acs{WAMR}, implementing an authority fence for single-writer cutovers across heterogeneous \acp{ISA} (§\ref{sec:methodology:prototype}).

We evaluate \ac{VSR} on a physical Raspberry Pi~5 testbed, through bidirectional x86-64/ARM64 migration trials, and under \acs{SGP4}-derived orbital schedules (§\ref{sec:evaluation}). Across 300 recurrent handovers, $p_{95}$ downtime remains below 6~ms without cumulative performance degradation, while all cross-\acs{ISA} trials resume successfully. Under realistic orbital profiles, migrations complete in 2.8--4.5~s between handovers spaced 83--242~s apart, yielding an availability of $A_d\ge99.997\%$.

\section{Motivation}
\label{sec:motivation}

Although \acp{GSL} between ground stations and satellites can achieve near-optimal propagation latencies of under $10~\text{ms}$~\cite{spacex_improving_nodate}, contemporary satellite networks operate primarily as terrestrial backhauls, introducing substantial routing overheads. This section analyzes the operational challenges of achieving \ac{VS} through an Industry 5.0 offshore wind farm maintenance use case.

\subsection{XR-Assisted Offshore Maintenance}
\label{sec:motivation:illustrative_scenario}

Consider an offshore wind farm operating beyond terrestrial edge connectivity. Technicians conducting turbine maintenance employ head-mounted \ac{XR} devices streaming real-time video, telemetry, and equipment state to an edge application providing interactive visual overlays and remote expert guidance~\cite{sharma_edge_2024,charpentier_paving_2024}. Sustaining these interactive workflows demands strict, ultra-low-latency execution; service degradation or latency jitter directly impairs task efficiency and operational safety. Hosting this application onboard passing \ac{LEO} satellites provides low-latency compute access without requiring capital-intensive offshore edge hardware. However, a typical maintenance window far exceeds the few-minute contact window of an individual satellite. Unassisted static deployment -- pinning the application to its initial host satellite -- causes the network path to stretch as the spacecraft moves past the horizon, necessitating a multi-hop inter-satellite relay which quickly causes latency \ac{SLO} violations.

\begin{figure}[h]
    \centering
    \resizebox{\linewidth}{!}{%
        \begin{tikzpicture}
\begin{axis} [
    only marks,
    width=\columnwidth,
    height=3.4cm,
    xlabel={Time (seconds)},
    ylabel={Latency (ms)},
    xlabel style={font=\small},
    ylabel style={font=\small},
    tick label style={font=\sffamily\small},
    grid=major,
    grid style={dashed, CustomAxes!15},
    unbounded coords=discard, 
    xmin=20, xmax=400,
    ymin=0, ymax=40,
    legend pos=north west,
]
    \addplot[
        mark=o, 
        color=CustomPrimary!90, 
        mark size=0.4pt,
    ] table [x index=0, y index=1, col sep=comma] {motivation/experiments/routing-latency/results_routing_latency.csv};
    
    \addplot[
        mark=o, 
        color=CustomPrimary!90, 
        mark size=0.4pt,
    ] table [x index=0, y index=2, col sep=comma] {motivation/experiments/routing-latency/results_routing_latency.csv};
    
    \addplot[
        mark=o, 
        color=CustomPrimary!90, 
        mark size=0.4pt,
    ] table [x index=0, y index=3, col sep=comma] {motivation/experiments/routing-latency/results_routing_latency.csv};
    

    \draw[dashed, CustomAlert!60!black] (axis cs:149, 0) -- (axis cs:149, 60) 
        node[pos=0.35, left, font=\footnotesize\color{CustomAlert!90!black}] {Handover};
        
    \draw[dashed, CustomAlert!60!black] (axis cs:269, 0) -- (axis cs:269, 60) 
        node[pos=0.45, left, font=\footnotesize\color{CustomAlert!90!black}] {Handover};

\end{axis}
\end{tikzpicture}%
    }
    \caption{End-to-end latency during satellite pass-over. Loss of direct line-of-sight and relay rerouting at 150~s and 270~s cause step-wise increases in baseline latency and jitter.}
    \label{fig:motivation:latency_of_not_moving_applications}
\end{figure}
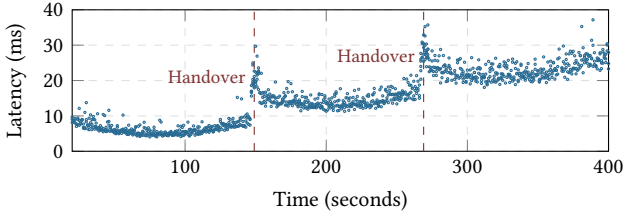

We empirically demonstrate this degradation in Figure~\ref{fig:motivation:latency_of_not_moving_applications}. During the initial direct line-of-sight pass-over, \ac{RTT} ranges between 4~ms and 11~ms ($p_5$ and $p_{95}$) due to dynamic slant distances and emulated network jitter. As orbital mechanics drive the host satellite below the horizon, traffic is rerouted over inter-satellite relays. This increases median \ac{RTT} from 6.6~ms ($\sigma = 2.4~\text{ms}$) to 14.4~ms ($\sigma = 2.9~\text{ms}$) for a single relay hop, reaching 23.1~ms ($\sigma = 3.3~\text{ms}$) across two hops within five minutes. Here, network \ac{RTT} alone exceeds 20~ms, exhausting the \ac{M2P} budget before sensing, processing, and rendering delays are included. This can induce a visual-vestibular misalignment, which increases the risk of simulator sickness that severely restricts technician operational time and safety~\cite{biswas_are_2024,palmisano_differences_2023,stauffert_effects_2018,hettinger_vection_1990}. Without migration, the worst-case co-planar route ($\ang{180}$ from the target) even requires 16 \ac{ISL} hops across a 550~km altitude shell. Extrapolating our measured $p_{95}$ penalty yields a propagation-only \ac{RTT} of 152.4~ms -- a delay that renders interactive \ac{XR} entirely unviable.

From the technician's perspective, the application should remain anchored directly above the wind farm throughout the maintenance session. Its logical location and active state must remain invariant despite the rapid physical movement of underlying spacecraft.

\subsection{Migrating the Application}
\label{sec:motivation:continuous_live_migration}

\begin{figure*}[!t]
    \centering

    \ref{legendmotivationcompare}

    \begin{subfigure}[t]{0.33\textwidth}
    \centering
    \begin{tikzpicture}
\begin{axis}[
    width=\linewidth,
    height=3.25cm,
    xlabel={Time (seconds)},
    ylabel={Packets Lost (\%)},
    xmin=10, xmax=30,
    ymin=0, ymax=120,
    yticklabel style={/pgf/number format/fixed, /pgf/number format/1000 sep={}},
    xlabel style={font=\small},
    ylabel style={font=\small},
    tick label style={font=\sffamily\small},
    set layers,
    grid=both,
    grid style={dashed, CustomAxes!15},
    legend style={
        at={(0.5, 1.4)},
        anchor=north,
        legend columns=3,
        draw=none,
        fill=none,
        /tikz/every even column/.append style={column sep=10pt}
    },
    legend to name=legendmotivationcompare,
]
    
    \begin{pgfonlayer}{axis background}
        \fill[CustomAlert!8] (axis cs:16,0) rectangle (axis cs:22,120);
        \draw[CustomAlert!60!black, thick, dashed] (axis cs:16,0) -- (axis cs:16,120);
        \draw[CustomAlert!60!black, thick, dashed] (axis cs:22,0) -- (axis cs:22,120);
    \end{pgfonlayer}

    \addlegendimage{area legend, fill=CustomAlert!8, draw=CustomAlert!60!black, thick, dashed}
    \addlegendentry{Handover Window}
    
    \node[anchor=center, text=CustomAlert!90!black, font=\sffamily\tiny\bfseries, align=center] 
        at (axis cs:22, 300) {Handover\\Window};
        
    \node[anchor=north, text=CustomPrimary, font=\sffamily\tiny\bfseries, align=center] 
        at (axis cs:11.5, 200) {Executing on\\Satellite 1};
        
    \node[anchor=north, text=CustomPrimary, font=\sffamily\tiny\bfseries, align=center] 
        at (axis cs:28.5, 200) {Executing on\\Satellite 2};

    
    \addplot[name path=upper, draw=none, forget plot] 
        table [col sep=comma, x=step, y=lost_p95] 
        {evaluation/experiments/packet-loss/packet_loss_results_stopcopy.csv};

    \addplot[name path=lower, draw=none, forget plot] 
        table [col sep=comma, x=step, y=lost_p5] 
        {evaluation/experiments/packet-loss/packet_loss_results_stopcopy.csv};

    \addplot[fill=CustomPrimary!50, draw=none, opacity=0.35] 
        fill between[of=upper and lower];
    \addlegendentry{$p_5\text{--}p_{95}$ Range}

    \addplot[
        color=CustomPrimary,
        line width=1.1pt,
    ] table [col sep=comma, x=step, y=lost_p50] 
      {evaluation/experiments/packet-loss/packet_loss_results_stopcopy.csv};
    \addlegendentry{$p_{50}$ Packet Loss (\%)}
    
    \node[
        anchor=north east,
        font=\sffamily\small,
        fill=white,
        draw=CustomAxes,
        rounded corners=4pt,
        inner sep=3pt
    ] at (rel axis cs:0.95,0.95) {a};
    
\end{axis}
\end{tikzpicture}%
    \phantomcaption
    \label{fig:motivation:sota_ill_suited_checkpoint_restore}
    \end{subfigure}
    \hfill
    \begin{subfigure}[t]{0.33\textwidth}
    \centering
    \begin{tikzpicture}
\begin{axis}[
    width=\linewidth,
    height=3.25cm,
    xlabel={Time (seconds)},
    ylabel={Packets Lost (\%)},
    xmin=10, xmax=35,
    ymin=0, ymax=50,
    log ticks with fixed point, 
    yticklabel style={/pgf/number format/fixed, /pgf/number format/1000 sep={}},
    xlabel style={font=\small},
    ylabel style={font=\small},
    tick label style={font=\sffamily\small},
    set layers,
    grid=both,
    grid style={dashed, CustomAxes!15},
    legend style={
        at={(0.5, 1.4)},
        anchor=north,
        legend columns=3,
        draw=none,
        fill=none,
        /tikz/every even column/.append style={column sep=10pt}
    },
    legend to name=legendmotivationcompare,
]
    
    \begin{pgfonlayer}{axis background}
        \fill[CustomAlert!8] (axis cs:18,0) rectangle (axis cs:30.5,50);
        \draw[CustomAlert!60!black, thick, dashed] (axis cs:18,0) -- (axis cs:18,50);
        \draw[CustomAlert!60!black, thick, dashed] (axis cs:30.5,0) -- (axis cs:30.5,50);
    \end{pgfonlayer}
    
    \addlegendimage{area legend, fill=CustomAlert!8, draw=CustomAlert!60!black, thick, dashed}
    \addlegendentry{Handover Window}
    
    \node[anchor=center, text=CustomAlert!90!black, font=\sffamily\tiny\bfseries, align=center] 
        at (axis cs:22, 300) {Handover\\Window};
        
    \node[anchor=north, text=CustomPrimary, font=\sffamily\tiny\bfseries, align=center] 
        at (axis cs:11.5, 200) {Executing on\\Satellite 1};
        
    \node[anchor=north, text=CustomPrimary, font=\sffamily\tiny\bfseries, align=center] 
        at (axis cs:28.5, 200) {Executing on\\Satellite 2};

    
    \addplot[name path=upper, draw=none, forget plot] 
        table [col sep=comma, x=step, y=lost_p95] 
        {evaluation/experiments/packet-loss/packet_loss_results_precopy.csv};

    \addplot[name path=lower, draw=none, forget plot] 
        table [col sep=comma, x=step, y=lost_p5] 
        {evaluation/experiments/packet-loss/packet_loss_results_precopy.csv};

    \addplot[fill=CustomPrimary!50, draw=none, opacity=0.35] 
        fill between[of=upper and lower];
    \addlegendentry{$p_5\text{--}p_{95}$ Range}

    \addplot[
        color=CustomPrimary,
        line width=1.1pt,
    ] table [col sep=comma, x=step, y=lost_p50] 
      {evaluation/experiments/packet-loss/packet_loss_results_precopy.csv};
    \addlegendentry{$p_{50}$ Packet Loss (\%)}
    
    \node[
        anchor=north east,
        font=\sffamily\small,
        fill=white,
        draw=CustomAxes,
        rounded corners=4pt,
        inner sep=3pt
    ] at (rel axis cs:0.95,0.95) {b};
    
\end{axis}
\end{tikzpicture}%
    \phantomcaption
    \label{fig:motivation:sota_ill_suited_pre_copy}
    \end{subfigure}
    \hfill
    \begin{subfigure}[t]{0.33\textwidth}
    \centering
    \begin{tikzpicture}
\begin{axis}[
    width=\linewidth,
    height=3.25cm,
    xlabel={Time (seconds)},
    ylabel={Packets Lost (\%)},
    xmin=10, xmax=40,
    ymin=0, ymax=120,
    log ticks with fixed point, 
    yticklabel style={/pgf/number format/fixed, /pgf/number format/1000 sep={}},
    xlabel style={font=\small},
    ylabel style={font=\small},
    tick label style={font=\sffamily\small},
    set layers,
    grid=both,
    grid style={dashed, CustomAxes!15},
    legend style={
        at={(0.5, 1.4)},
        anchor=north,
        legend columns=3,
        draw=none,
        fill=none,
        /tikz/every even column/.append style={column sep=10pt}
    },
    legend to name=legendmotivationcompare,
]
    
    \begin{pgfonlayer}{axis background}
        \fill[CustomAlert!8] (axis cs:16,0) rectangle (axis cs:35,120);
        \draw[CustomAlert!60!black, thick, dashed] (axis cs:16,0) -- (axis cs:16,120);
        \draw[CustomAlert!60!black, thick, dashed] (axis cs:35,0) -- (axis cs:35,120);
    \end{pgfonlayer}

    \addlegendimage{area legend, fill=CustomAlert!8, draw=CustomAlert!60!black, thick, dashed}
    \addlegendentry{Handover Window}
    
    \node[anchor=center, text=CustomAlert!90!black, font=\sffamily\tiny\bfseries, align=center] 
        at (axis cs:22, 300) {Handover\\Window};
        
    \node[anchor=north, text=CustomPrimary, font=\sffamily\tiny\bfseries, align=center] 
        at (axis cs:11.5, 200) {Executing on\\Satellite 1};
        
    \node[anchor=north, text=CustomPrimary, font=\sffamily\tiny\bfseries, align=center] 
        at (axis cs:28.5, 200) {Executing on\\Satellite 2};

    
    \addplot[name path=upper, draw=none, forget plot] 
        table [col sep=comma, x=step, y=lost_p95] 
        {evaluation/experiments/packet-loss/packet_loss_results_postcopy.csv};

    \addplot[name path=lower, draw=none, forget plot] 
        table [col sep=comma, x=step, y=lost_p5] 
        {evaluation/experiments/packet-loss/packet_loss_results_postcopy.csv};

    \addplot[fill=CustomPrimary!50, draw=none, opacity=0.35] 
        fill between[of=upper and lower];
    \addlegendentry{$p_5\text{--}p_{95}$ Range}

    \addplot[
        color=CustomPrimary,
        line width=1.1pt,
    ] table [col sep=comma, x=step, y=lost_p50] 
      {evaluation/experiments/packet-loss/packet_loss_results_postcopy.csv};
    \addlegendentry{$p_{50}$ Packet Loss (\%)}
    
    \node[
        anchor=north east,
        font=\sffamily\small,
        fill=white,
        draw=CustomAxes,
        rounded corners=4pt,
        inner sep=3pt
    ] at (rel axis cs:0.95,0.95) {c};
    
\end{axis}
\end{tikzpicture}%
    \phantomcaption
    \label{fig:motivation:sota_ill_suited_post_copy}
    \end{subfigure}

    \caption{Packet loss during inter-satellite handover of a 256~MB UDP-based \ac{XR} workload. Red shading marks handover windows; lines and bands show the median and p5-p95 range. (\subref{fig:motivation:sota_ill_suited_checkpoint_restore})~Checkpoint/restore causes a complete blackout. (\subref{fig:motivation:sota_ill_suited_pre_copy})~Pre-copy limits loss but prolongs handover. (\subref{fig:motivation:sota_ill_suited_post_copy})~Post-copy incurs sustained loss from reactive page faults.}
    \label{fig:motivation:sota_ill_suited}
\end{figure*}
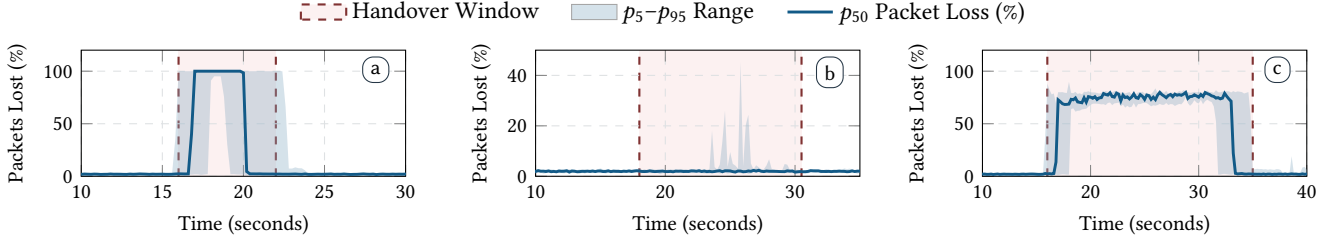

Continuous, high-frequency migration converts live migration from an exception-handling mechanism into a core steady-state execution primitive. To evaluate migration under this assumption, we extended \acs{WAMR} to support three canonical migration paradigms. We omit kernel tools such as \acs{CRIU} as they require OS configurations that are not supported on constrained edge hardware~\cite{noauthor_criu_2021}. We deployed the experiment on a two-node Raspberry~Pi~5 testbed, emulating satellite nodes receiving UDP telemetry from a \textit{VR Gaming Dataset} trace ($\approx 4{,}000$~pkts/s; $\approx 5~\text{MB/s}$)~\cite{michaelides_vr_2023}. Figure~\ref{fig:motivation:sota_ill_suited} illustrates the performance trade-offs of these mechanisms during handover.

\textit{Checkpoint/Restore} halts execution on the source node, serializes the complete memory and CPU register state into a snapshot, transfers it across the network, and restores execution on the target host~\cite{chen_checkpoint_2015,bozyigit_user-level_2001}. Despite its conceptual simplicity, Figure~\ref{fig:motivation:sota_ill_suited_checkpoint_restore} shows that completely pausing application execution during state transfer induces a full service blackout with $100\%$ packet loss. \textit{Pre-Copy} transfers an initial memory snapshot to the target before repeatedly transmitting dirtied memory pages while the application continues to execute on the source~\cite{theimer_preemptable_1985,govindaraj_container_2018}. While this strategy maintains low baseline packet loss (Figure~\ref{fig:motivation:sota_ill_suited_pre_copy}), high memory write rates can prevent iterative convergence~\cite{clark_live_2005}. Consequently, the overall migration duration inflates past 10~s, consuming a prohibitive fraction of the 15~s beam-switch window typical for Starlink~\cite{zhao_demystifying_2026}. \textit{Post-Copy} pauses the source instance only long enough to transfer minimal execution registers, immediately spawning the process on the target host without its memory. As execution resumes, missing memory pages are fetched reactively from the source host~\cite{hines_post-copy_2009,puliafito_container_2019}. Although initial downtime remains negligible, Figure~\ref{fig:motivation:sota_ill_suited_post_copy} demonstrates that fetching unmapped memory on demand across the inter-satellite link triggers severe latency jitter and sustained packet loss of $\approx 75\%$.

Given the brief contact windows and the compute and energy limits of satellites, traditional terrestrial trade-offs that sacrifice migration time for downtime, or vice versa, are prohibitive.

\section{Related Work}
\label{sec:related_work}

Stateful migrations are commonly implemented as individually triggered infrastructure operations~\cite{attar-khorasani_seamless_2025,soussi_democratizing_2024,kaur_container_2022}. 
ReSync~\cite{scheuer_resync_2026} couples \acs{CRIU} checkpoint/restore with buffered application-input replay and coordinates migration with mobile-network handover. Instead, \ac{VSR} leaves placement and timing to an external orchestrator, and projects observed host-call returns and application-memory effects. The successor can therefore reconstruct execution without repeating external effects. Voyager~\cite{nadgowda_voyager_2017} combines \acs{CRIU} memory transfer with a networked union filesystem, from which the target retrieves persistent state during lazy transfer. \ac{VSR} does not federate host resources or require predecessor access after commit. It rebinds logical capabilities to target-local resources. dOS/RecShim~\cite{bergan_deterministic_2010} enforces deterministic process groups and records external nondeterminism for replay and replication. \ac{VSR} applies the same separation between deterministic execution and external effects to project a live computation across successive authoritative hosts rather than provide general deterministic execution. ReSync, Voyager, and dOS/RecShim rely on OS- or kernel-level mechanisms and generally reproduce the same host environment at the destination. \ac{VSR} transfers portable runtime state and rebinds declared capabilities, avoiding common-kernel and common-\acs{ISA} requirements across satellite compute platforms with heterogeneous processors and payload capabilities.

ReVirt~\cite{dunlap_revirt_2002} logs nondeterministic events beneath a guest OS for instruction-level VM replay and post-intrusion analysis. Liu et al.~\cite{liu_live_2009} combine a copy-on-write VM checkpoint with iterative transfer and target-side full-system replay. Their design is mechanically close to continuous state projection, but runs at the VM boundary and targets a single handover. Yu et al.~\cite{yu_live_2015} adapt iterative logging and replay to changes above a shared Docker image. \ac{VSR} instead defines a runtime protocol in which boundary events record host-call results and explicit application-memory mutations, replay suppresses new external effects, capabilities are rebound at the successor, and an authority fence prevents duplicate effects.

Bhattacherjee et al.~\cite{bhattacherjee_-orbit_2020} introduced \ac{VS} and evaluated placement policies that keep a service near a fixed ground location. Wu et al.~\cite{wu_task_offloading_service_migration_2024} jointly optimize task offloading and service migration, while Databelt~\cite{marcelino_databelt_2025} constructs a data path for orbital serverless workflows. Komet~\cite{pfandzelter_komet_2024} repeatedly instantiates stateless functions and proactively replicates application-managed state. These systems select or prepare the next placement but do not specify how to transfer a running execution, re-establish its capabilities, and preserve a single authority across recurrent handovers. Orbital handover windows are bounded and may terminate access to the predecessor. \ac{VSR} complements this placement work by accepting the successor and constraints from an external orchestrator, terminating predecessor dependence at commit, and exposing downtime, migration duration, and long-run migration occupancy for feasibility analysis.

\section{A Runtime for Virtual Stationarity}
\label{sec:methodology}

We first formalize the correctness and feasibility contract for recurrent migration (\S\ref{sec:methodology:formalization}). We then present the \acf{VSR}, separating its runtime responsibilities from the external orchestration and execution-engine services it assumes (\S\ref{sec:methodology:framework}). Finally, we detail the membrane's two enabling mechanisms (\S\ref{sec:methodology:membrane}) and report one concrete realization (\S\ref{sec:methodology:prototype}).

\subsection{System and Feasibility Model}
\label{sec:methodology:formalization}

We model an application anchored to a ground target $g$ over an operational interval $\mathcal{T}$, with duration $|\mathcal{T}|$. The application may execute on any satellite $S_i$ in the constellation. Its end-to-end \ac{RTT} to $S_i$ at time $t$ is described by the latency process $\zeta_i: \mathcal{T}\rightarrow\mathbb{R}^{+}\cup\{\infty\}$, where $\zeta_i(t)=\infty$ denotes that no \ac{GSL}/\ac{ISL} path exists. Orbital motion makes the geometric component of $\zeta_i$ predictable, while atmospheric, processing, and queueing delays remain time-varying.

\subsubsection{Continuity Contract}
\label{sec:methodology:formalization:contract}

Let $\sigma(t)=\langle E(t),M(t),C(t),g\rangle$ denote the logical application state: execution context $E$, mutable state $M$, logical capability contract $C$, and geographic anchor $g$. Host $S_i$ realizes this logical state as $\rho_i(t)=\pi_i(\sigma(t))$ by translating execution and binding logical capabilities to native implementations. Two realizations are equivalent if, for the same subsequent inputs, they produce the same application-visible state and external effects.

A correct handover preserves four invariants. \emph{State continuity} requires the target to continue from an equivalent application-visible state. \emph{Capability continuity} preserves the identity and semantics of functions classified as required when the successor is selected; the availability of optional functions may differ across hosts and is explicitly observable by the application. \emph{Effect continuity} prevents externally visible operations from being omitted or repeated during reconstruction. Finally, \emph{authority exclusivity} permits only one realization to produce new external effects. A successor may consequently be reconstructed while the source runs, but it remains non-authoritative until handover completes.

\subsubsection{Service Objectives and Hosting Windows}

The application specifies a latency \ac{SLO} $S_{max}$ and an availability \ac{SLO} $A_{min}$. Let $D\subseteq\mathcal{T}$ contain all times at which the application does not answer within $S_{max}$, including execution pauses, path-update delays, and ordinary latency violations. End-to-end availability is therefore

\begin{equation}
    A = 1 - \frac{|D|}{|\mathcal{T}|} \geq A_{min}.
    \label{eq:availability}
\end{equation}

A satellite is an admissible host only while its latency remains within the \ac{SLO}. Starting at time $t$, its remaining hosting window is

\begin{equation}
    T_{host}(S_i,t) =
    \sup\!\left\{\Delta t\geq0 \;\middle|\;
    \forall \tau\in[t,t+\Delta t]:\zeta_i(\tau)\leq S_{max}\right\}.
    \label{eq:hosting_window}
\end{equation}

Consider handover $k$ from source $S_i$ to successor $S_j$, triggered at $t_k$. The remaining source window provides the migration budget $B_k=T_{host}(S_i,t_k)$. A migration mechanism produces a completion time $t_{mig,k}$ and an execution interruption $d_k$. To complete the handover safely, it must satisfy $t_{mig,k}\leq B_k$, and the selected successor must be \ac{SLO}-compliant when authority transfers.

\subsubsection{Steady-State Feasibility}

Because migration recurs throughout $\mathcal{T}$, its cost must be evaluated cumulatively rather than as an isolated event. For $n$ handovers, the fraction of operational time occupied by migration is

\begin{equation}
    \phi_{mig} = \frac{\sum_{k=1}^{n} t_{mig,k}}{|\mathcal{T}|}
    \approx \frac{\overline{t}_{mig}}{\overline{T}_{host}},
    \label{eq:migration_overhead}
\end{equation}

where the approximation applies when each hosting window induces one handover. Migration may overlap application execution, so $\phi_{mig}$ captures sustained link and compute occupancy rather than service downtime. If only execution pauses are considered, they yield the diagnostic \mbox{$A_d=1-\sum_k d_k/|\mathcal{T}|$}; $A_d$ equals the availability in Equation~\ref{eq:availability} only when other latency violations are absent.

\subsubsection{Contract}
Together, the continuity invariants and operational constraints define the mechanism-independent \ac{VSR} contract. Each migration must complete within its budget and commit to an \ac{SLO}-compliant successor. Across the execution schedule, availability must remain above $A_{min}$ and migration occupancy below the application-defined limit $\epsilon$. Accordingly, feasibility requires

\begin{equation}
\boxed{
\begin{aligned}
    \forall k:\quad
    &t_{mig,k} \leq B_k, \\
    &\zeta_j(t_k+t_{mig,k}) \leq S_{max},\\
    &A \geq A_{min}, \\
    &\phi_{mig} \leq \epsilon.
\end{aligned}
}
\label{eq:vs_feasibility}
\end{equation}

\subsection{Virtual Stationarity Runtime Framework}
\label{sec:methodology:framework}

\begin{figure}[h]
    \centering
    \includegraphics[width=\linewidth]{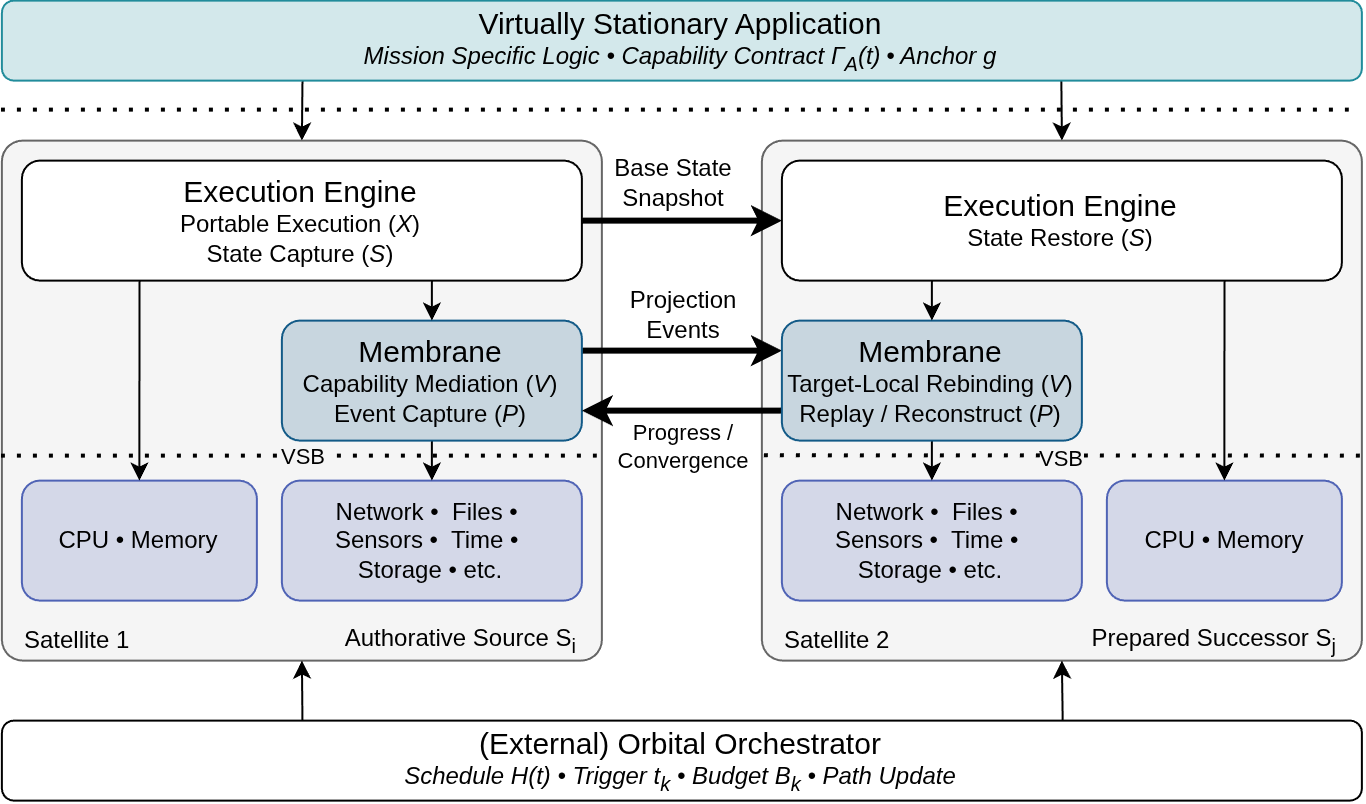}
    \caption{Virtual stationarity runtime. External orchestration supplies the selected host and trigger. The Execution Engine supplies portable execution and state capture; the Migration Membrane preserves capabilities, projects continued execution, and controls authority transfer.}
    \label{fig:methodology:architecture}
\end{figure}

The model specifies what a handover must achieve, but not which component is responsible for each obligation. As shown in Figure~\ref{fig:methodology:architecture}, the framework relates an external orbital orchestrator, a compatible Execution Engine, and the Migration Membrane. It deliberately separates the external decision of \emph{when and where} to migrate from the runtime mechanisms that determine \emph{how} execution continues.

\subsubsection{Components}
Our framework distinguishes three core components to enable recurrent migrations.

\emph{External Orbital Orchestrator.}
The orchestrator predicts host $\mathcal{H}(t)$ and network evolution, applies Equation~\ref{eq:vs_feasibility} and Equation~\ref{eq:capability_compatibility}, and supplies successor $S_j$, trigger $t_k$, and migration budget $B_k$. It thereby defines the admissible handover, while constellation prediction, host-selection policy, and routing remain outside the \ac{VSR}. The runtime contributes to Equation~\ref{eq:vs_feasibility} by reducing migration time $t_{mig,k}$ and interruption $d_k$ within the supplied budget.

\emph{Execution Engine.}
The framework does not require a particular bytecode or managed runtime, but the engine must provide two properties. \emph{Portable execution} ($\mathcal{X}$) represents code and control state independently of the source OS and \ac{ISA}. \emph{State capture and restoration} ($\mathcal{S}$) exposes safe points and a complete, relocatable representation of the state required to continue $\mathcal{X}$. The engine also exposes the application-memory mutations caused by an intercepted system call. These are integration requirements.

\emph{Migration Membrane.}
The membrane binds to the engine at the \acf{VSB}, separating logical execution from native resources and effects. It contributes Capability Virtualization ($\mathcal{V}$) and Continuous State Projection ($\mathcal{P}$), coordinates the handover protocol, and enforces authority exclusivity. Each realization operates in one of four modes: \textsc{Live} performs normal authoritative execution; \textsc{Record} performs authoritative execution while emitting continuation events; \textsc{Replay} reconstructs those events without external effects; and \textsc{Inactive} holds prepared bindings that cannot be invoked.

\subsubsection{Handover Protocol}
\label{sec:methodology:framework:lifecycle}
Our framework structures recurrent migration into four distinct phases.

\emph{Prepare.}
The membrane validates the supplied migration parameters and requests a source safe point. With no system call in flight, $\mathcal{S}$ captures the portable base $S_{base}$. The source initializes its opcode counter relative to this cut; the target initializes the corresponding counter after restoring the same base.

\emph{Reconstruct.}
The base is transferred while the source continues in \textsc{Record}. The target restores $S_{base}$, starts a receiver for continuation events, and uses $\mathcal{V}$ to establish target-local capability bindings. These bindings remain \textsc{Inactive} and thus cannot produce effects.

\emph{Project.}
After restoration, the target enters \textsc{Replay}. The source continues intercepting system calls and emitting their results and memory mutations; the target consumes the events and reconstructs the same transitions without invoking its local bindings. Projection continues until replay lag satisfies the commit threshold.

\emph{Commit.}
The source quiesces at an opcode boundary, closes its effect gate, transitions from \textsc{Record} to \textsc{Inactive}, and emits a terminal \textsc{End} event containing its execution fence. After consuming all preceding events and reaching the same fence, the target activates its bindings and transitions from Replay to Live. The runtime then requests the orchestrator’s path update.

\subsection{Migration Membrane Mechanisms}
\label{sec:methodology:membrane}

A portable snapshot can reconstruct a stopped computation, but it neither preserves the meaning of host-bound resources nor captures state accumulated while the snapshot is in transit. The Migration Membrane $\mathcal{M}= \left\langle \mathcal{V},\mathcal{P}\right\rangle$ addresses both gaps, where $\mathcal{V}$ mediates the application's relationship with its physical environment and $\mathcal{P}$ maintains a non-authoritative projection of its evolving execution.

\subsubsection{Capability Virtualization ($\mathcal{V}$)}
\label{sec:methodology:membrane:capabilities}

Application $A$ expresses each host-dependent operation as a logical capability $c\in C_A$ with a stable name independent of any native handle or implementation. Each satellite $S_i$ has a static inventory $\mathcal{C}_i$ of capabilities it can implement. The runtime uses the same logical namespace on every host and changes only the local binding $\beta_i(c)$.

\textit{Declaration.}
The application capability contract is
\begin{equation}
    \Gamma_A(t)=\langle C_A^{req}(t),C_A^{opt}(t)\rangle,
    \qquad
    C_A^{req}(t)\mathbin{\dot\cup}C_A^{opt}(t)=C_A,
    \label{eq:capability_contract}
\end{equation}
The initial capability set can be declared explicitly or extracted by inspecting the application's source-level declarations, intermediate representation, or compiled external-call interface. Capabilities are required by default. During execution, the application may call $\mathsf{SetPolicy}(c,p)$, with $p\in\{\mathsf{required},\mathsf{optional}\}$, to update the contract stored in logical state $C$. Policy changes are deterministic control operations applied by both realizations and require no distinct continuation-event type. For optional capabilities, the membrane exposes a function so the application can adapt its behavior according to the active host.

\begin{equation}
    \mathsf{Available}(c)=
    \begin{cases}
        \mathsf{true}, & c\in\mathcal{C}_{\mathcal{H}(t)},\\
        \mathsf{false}, & \text{otherwise},
    \end{cases}
    \label{eq:capability_available}
\end{equation}

\emph{Admission.}
The orbital orchestrator admits only candidates satisfying

\begin{equation}
    \mathsf{Compatible}(A,S_i,t)
    \iff C_A^{req}(t)\subseteq\mathcal{C}_i.
    \label{eq:capability_compatibility}
\end{equation}

Required capabilities therefore constrain placement in addition to Equation~\ref{eq:vs_feasibility}. Optional capabilities do not constrain placement; they constrain application behavior through Equation~\ref{eq:capability_available}. The contract is sampled when the successor is selected, and later policy changes affect the next placement decision.

\emph{Preparation.}
During reconstruction, the target evaluates $\beta_j(c)\leftarrow\mathsf{Rebind}_j(c)$ for every capability. Available capabilities obtain host-local resources; a missing optional capability maps to the defined \textsc{Unavailable} behavior, whereas a missing required capability invalidates admission. Preparation establishes native resources but leaves all target bindings \textsc{Inactive}; it grants neither execution authority nor permission to produce external effects.

\emph{Activation.}
While the source remains authoritative, system calls invoke $\beta_i(c)$ in \textsc{Live} or \textsc{Record}. The target in \textsc{Replay} returns recorded outcomes without invoking $\beta_j(c)$. Only after the handover protocol commits are the source bindings revoked and the target bindings activated, so subsequent calls use target-local implementations. Consequently, logical capability identity persists across hosts while only the authoritative realization may produce new external effects.

\subsubsection{Continuous State Projection ($\mathcal{P}$)}
\label{sec:methodology:membrane:projection}

Continuous State Projection reconstructs a target from a portable base and an ordered continuation while the source remains authoritative. We assume reliable FIFO channels and single-threaded modules whose host interactions occur synchronously through the VSB. Execution between VSB crossings is deterministic.

\emph{State capture.}
At trigger $t_0$, the source reaches a safe point with no system call in flight. $\mathcal{S}$ captures $S_{base}=\mathsf{Capture}(\sigma(t_0))$, while paused, and the source initializes opcode counter $q_s=0$. The source realization then enters \textsc{Record} before resuming. The target initializes $q_t=0$ after restoring the same cut. The execution-engine contract includes mutation tracking for intercepted system calls. Before invoking a native operation, the tracker is reset. Runtime paths that permit the operation to modify application memory register each affected pointer and length. Immediately after the call returns, a collect-and-reset operation converts those pointers to relocatable memory offsets and copies the resulting bytes:

\begin{equation}
    \Delta M_k=
    \{\langle o_{k,\ell},n_{k,\ell},b_{k,\ell}\rangle\}_{\ell=1}^{m_k},
    \label{eq:memory_mutations}
\end{equation}

where $o_{k,\ell}$ is an offset, $n_{k,\ell}$ the region length, and $b_{k,\ell}$ the bytes visible after the call. An implementation may expose this as an engine API or as an internal additional return value of native invocation. The continuation event is

\begin{equation}
    e_k=\langle k,s_k,r_k,\Delta M_k\rangle,
    \qquad s_k=\mathsf{OpId}(c_k,f_k),
    \label{eq:continuation_event}
\end{equation}

\begin{algorithm}[t]
\caption{VSB-mediated capability invocation}
\label{alg:vsb_invocation}
\begin{algorithmic}[1]
\Require Capability $c$, operation $f$, arguments $a$, mode
         $m\in\{\mathsf{Live},\mathsf{Record},\mathsf{Replay},
         \mathsf{Inactive}\}$, binding $\beta_i$, target FIFO $Q$
\If{$m=\mathsf{Replay}$}
    \State $e\gets Q.\mathsf{DequeueCallBlocking}()$
    \If{$e=\mathsf{finalized}\lor e.s\neq\mathsf{OpId}(c,f)$}
        \State \Return \textsc{Abort}
    \EndIf
    \State $\mathsf{Apply}(e.\Delta M)$; $h_t\gets h_t+1$
    \State \Return $e.r$ without invoking $\beta_i(c)$
\EndIf
\If{$m=\mathsf{Inactive}$}
    \State \Return \textsc{Abort}
\EndIf
\If{$\beta_i(c)=\bot$}
    \State $r\gets\mathsf{Unavailable}$; $\Delta M\gets\emptyset$
\ElsIf{$m=\mathsf{Record}$}
    \State $(r,\Delta M)\gets
        \mathsf{InvokeAndCapture}(\beta_i(c),f,a)$
\Else
    \State $r\gets\mathsf{Invoke}(\beta_i(c),f,a)$
\EndIf
\If{$m=\mathsf{Record}$}
    \State $h_s\gets h_s+1$
    \State $\mathsf{Emit}(\langle h_s,\mathsf{OpId}(c,f),r,\Delta M\rangle)$
\EndIf
\State \Return $r$
\end{algorithmic}
\end{algorithm}

where $k$ is the FIFO event counter, $s_k$ the operation signature, and $r_k$ is the direct return value. Separating $r_k$ from $\Delta M_k$ is essential for reads: a file or socket read may return only a status or byte count in $r_k$, while the received content resides in the memory regions of $\Delta M_k$. The source enqueues $e_k$ before returning to the application, and an asynchronous sender preserves event order.

\emph{Transfer and replay.}
Base state and continuation events use independent channels. A receiver places events into the ordered buffer $\mathcal{Q}$, so replay never reads from the transport directly. Algorithm~\ref{alg:vsb_invocation} applies each event without invoking the target binding; mismatches or buffer overflow abort reconstruction while the source remains authoritative. The terminal \textsc{End} event is handled separately as the authority fence.

\emph{Convergence.}
Let $h_s$ be the number of emitted events and $h_t$ the number acknowledged after replay. Their difference is the observable lag $L=h_s-h_t$. Then $T_0=t_{ovh}+\Phi/C$ denotes the time to capture, transfer, and restore a base snapshot of size $\Phi$ at effective bandwidth $C$. Events accumulate at rate $r_{sys}$, giving an initial replay backlog $Q_0=r_{sys}T_0$. While the source continues, a target processing events at rate $\mu$ drains this backlog at rate $\mu-r_{sys}$. Hence

\begin{equation}
    t_{mig}^{\mathcal{M}} =
    T_0 + \frac{Q_0}{\mu-r_{sys}}
    = \left(t_{ovh} + \frac{\Phi}{C}\right)
      \frac{\mu}{\mu-r_{sys}},
    \qquad r_{sys}<\mu.
    \label{eq:migration_time}
\end{equation}

Convergence requires $r_{sys}<\mu$. For $r_{sys}\ll\mu$, migration time is approximately linear in $\Phi$ at slope $1/C$ and independent of memory-dirty rate. As $r_{sys}$ approaches $\mu$, the continuation backlog ceases to converge. Pre-copy instead approaches $t_{pre}\approx(\Phi/C)/(1-r_{dirty}/C)$ and diverges as $r_{dirty}$ approaches $C$~\cite{clark_live_2005}. If $r_{sys}\geq\mu$ or predicted $t_{mig,k}^{\mathcal{M}}>B_k$, the framework aborts target reconstruction and leaves the source authoritative.

\emph{Authority transfer.}
The condition $L\leq L_{conv}$ only triggers commit; it does not prove that source and target are at the same instruction state. Both realizations therefore maintain the opcode counters initialized at the base cut and increment them after every completed opcode. At commit, the source quiesces at an opcode boundary and fixes fence $\langle q^\star,\iota^\star\rangle$, where $q^\star$ is its completed-opcode count and $\iota^\star$ identifies the opcode at that boundary. It then changes from \textsc{Record} to \textsc{Inactive}, closing the source effect gate before any target activation. The sender flushes all ordinary events and appends the final packet

\begin{equation}
    e_{end}=\langle\mathsf{END},q^\star,\iota^\star\rangle.
    \label{eq:authority_fence}
\end{equation}

FIFO order makes \textsc{End} an explicit statement that no further source-produced system-call outcome can follow. The target receiver stores the fence separately and marks $Q$ finalized after all preceding events have been inserted. The target remains in \textsc{Replay}, applies the remaining events, and checks $\langle q_t,\iota_t\rangle$ after each opcode. It transitions from \textsc{Replay} to \textsc{Live} and activates $\beta_j$ only when $Q$ is empty and $\langle q_t,\iota_t\rangle=\langle q^\star,\iota^\star\rangle$. If the target passes the counter or the opcode identifier differs, commit fails without enabling effects. Thus, source revocation precedes target activation and there is never more than one effect-producing realization. Stream closure remains a transport indication; Equation~\ref{eq:authority_fence} identifies the exact execution cut. After activation, the runtime requests the orchestrator's path update. Failures before source revocation abort the projection; recovery after revocation is outside scope.

\subsection{Prototype Implementation}
\label{sec:methodology:prototype}

We integrate our approach with the \ac{WAMR}. An interpreter hook after each opcode checks whether migration has been triggered. At the selected boundary, the interpreter thread captures application state together with stack and shadow-stack frames. To minimize this pause, it creates a read-only \texttt{MAP\_PRIVATE} view of the linear-memory \texttt{memfd} backing object~\cite{kerrisk_memfd_create2_2026}. Copy-on-write preserves the base while the source resumes and the image is transmitted. The target restores the relocatable structures, providing $\mathcal{X}$ and $\mathcal{S}$ across operating systems and \acp{ISA}.

We adapt \ac{WAMR}'s native-import boundary to realize $\mathcal{V}$. At module load, the runtime derives logical capabilities from imported module/function pairs. Missing optional \acs{WASI} functions return \texttt{\_\_WASI\_ERRNO\_ENOSYS}. The evaluated UDP binding captures its descriptor slot and port, creates a target socket, and updates \ac{WAMR}'s descriptor table to the new native socket.

For $\mathcal{P}$, the mutation tracker is reset before native dispatch and records the linear-memory regions written by the system call. After dispatch, the runtime serializes each record as offset, length, and payload alongside a function-signature hash and two return cells. The execution thread appends these events to ping-pong buffers; a flusher thread swaps them at most every $500\,\mu$s and transmits completed batches without blocking execution on network I/O. On the target, a receiver thread inserts events into a bounded FIFO buffer consumed by Algorithm~\ref{alg:vsb_invocation}, while periodic acknowledgements expose $h_t$ to the source. At commit, the source stops at the next opcode, flushes the final batch, and appends \textsc{End} with the source opcode counter and identifier. The target opcode hook activates execution only at the matching local tuple. The prototype currently supports single-threaded modules; the evaluated calls modify one contiguous output region. Anonymized source code and build scripts for our adapted WAMR runtime prototype are available for review\footnote{\url{https://github.com/danielkleebinder/migration-membrane}}.
\section{Evaluation}
\label{sec:evaluation}

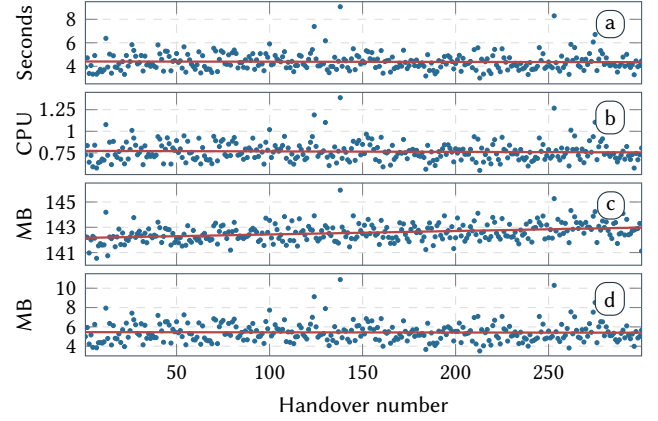
\begin{figure}[t]
    \centering
    \resizebox{\linewidth}{!}{%
        \begin{tikzpicture}
\begin{groupplot}[
    group style={
        group size=1 by 4,
        vertical sep=3pt
    },
    width=\linewidth,
    height=2.6cm,
    xmin=1,
    xmax=300,
    xtick={50,100,150,200,250},
    grid=major,
    grid style={dashed, CustomAxes!15},
    axis line style={CustomAxes!80},
    scaled ticks=false,
    clip marker paths=true,
    xlabel style={
        font=\sffamily\small
    },
    ylabel style={
        rotate=0,
        at={(axis description cs:-0.11,0.5)},
        anchor=center,
        font=\sffamily\small
    },
    tick label style={
        font=\sffamily\small
    },
    legend style={
        at={(0.02, 0.95)},
        anchor=north west,
        font=\sffamily\scriptsize,                  
        fill=white,
        fill opacity=0.85,
        draw=CustomAxes!30,
        thin,
        row sep=-1pt,
        legend columns=1,
        legend cell align={left},                    
        /tikz/every odd column/.append style={column sep=8pt},   
        /tikz/every even column/.append style={column sep=10pt} 
    },
]

\nextgroupplot[
    xlabel={},
    xticklabels={},
    ylabel={Seconds},
    ymin=2.5,
    ymax=9.5,
    ytick={4,6,8}
]

\addplot[
    only marks,
    mark=*, 
    color=CustomPrimary!90, 
    mark size=0.7pt,
] table[
    col sep=comma,
    x=HandoverId,
    y=MigrationTimeS
] {evaluation/experiments/recurrent-migration/stats_continuous_handover.csv};

\addplot[
    CustomAlert!85!black,
    thick
] coordinates {
    (1,   4.40577242)
    (300, 4.34993272)
};

\node[
    anchor=north east,
    font=\sffamily\small,
    fill=white,
    draw=CustomAxes,
    rounded corners=4pt,
    inner sep=3pt
] at (rel axis cs:0.97,0.9) {a};

\nextgroupplot[
    xlabel={},
    xticklabels={},
    ylabel={CPU},
    ymin=0.5,
    ymax=1.45,
    ytick={0.75, 1, 1.25}
]

\addplot[
    only marks,
    mark=*, 
    color=CustomPrimary!90, 
    mark size=0.7pt,
    forget plot
] table[
    col sep=comma,
    x=HandoverId,
    y=CpuCoreSeconds
] {evaluation/experiments/recurrent-migration/stats_continuous_handover.csv};

\addplot[
    CustomAlert!85!black,
    thick,
    forget plot
] coordinates {
    (1,   0.77335139)
    (300, 0.75618456)
};

\node[
    anchor=north east,
    font=\sffamily\small,
    fill=white,
    draw=CustomAxes,
    rounded corners=4pt,
    inner sep=3pt
] at (rel axis cs:0.97,0.9) {b};

\nextgroupplot[
    xlabel={},
    xticklabels={},
    ylabel={MB},
    ymin=140,
    ymax=146.5,
    ytick={141,143,145}
]

\addplot[
    only marks,
    mark=*, 
    color=CustomPrimary!90, 
    mark size=0.7pt,
    forget plot
] table[
    col sep=comma,
    x=HandoverId,
    y=PostCleanupMemoryMB
] {evaluation/experiments/recurrent-migration/stats_continuous_handover.csv};

\addplot[
    CustomAlert!85!black,
    thick,
    forget plot
] coordinates {
    (1,   142.16205755)
    (300, 142.97861953)
};

\node[
    anchor=north east,
    font=\sffamily\small,
    fill=white,
    draw=CustomAxes,
    rounded corners=4pt,
    inner sep=3pt
] at (rel axis cs:0.97,0.9) {c};

\nextgroupplot[
    xlabel={Handover number},
    ylabel={MB},
    ymin=3,
    ymax=11.5,
    ytick={4,6,8,10}
]

\addplot[
    only marks,
    mark=*, 
    color=CustomPrimary!90, 
    mark size=0.7pt,
    forget plot
] table[
    col sep=comma,
    x=HandoverId,
    y=EventTransferMB
] {evaluation/experiments/recurrent-migration/stats_continuous_handover.csv};

\addplot[
    CustomAlert!85!black,
    thick,
    forget plot
] coordinates {
    (1,   5.45483643)
    (300, 5.39853369)
};

\node[
    anchor=north east,
    font=\sffamily\small,
    fill=white,
    draw=CustomAxes,
    rounded corners=4pt,
    inner sep=3pt
] at (rel axis cs:0.97,0.9) {d};

\end{groupplot}
\end{tikzpicture}%
    }
    \caption{Stability across 300 accelerated handovers with a 30\,s trigger interval. Points report (a) per-handover migration time, (b) aggregate CPU consumption across both hosts, (c) post-cleanup memory, and (d) transferred continuation events. Lines show least-squares trends.}
    \label{fig:evaluation:recurrent_migration}
\end{figure}

We evaluate our prototype on a physical testbed of four Raspberry Pi~5 \acp{SBC} connected by an unmanaged Gigabit switch. Each node has a quad-core 2.4~GHz ARMv8 processor, 8~GB of RAM, an active cooling unit, and runs Raspberry Pi OS (Debian~13). Raspberry Pi devices represent \ac{COTS} payload classes rather than specific flight hardware. Devices from these families have flown on satellites before~\cite{xing_deciphering_2024,furutanpey_fool_2025,cratere_-board_2024}. The evaluation follows our three contributions: §\ref{sec:evaluation:stability} and §\ref{sec:evaluation:continuity} test steady-state execution and semantic continuity, §\ref{sec:evaluation:scaling_and_convergence_bounds} quantifies the projection bounds in Equation~\ref{eq:migration_time}, and §\ref{sec:evaluation:portability} and §\ref{sec:evaluation:orbital_feasibility} evaluate cross-\ac{ISA} execution and orbital feasibility using $t_{mig}$, $d$, $\phi_{mig}$, and $A_d$. Because path updates are external to \ac{VSR}, $A_d$ measures runtime interruption rather than the end-to-end availability $A$ in Equation~\ref{eq:availability}.

\subsection{Steady-State Correctness and Stability}
\label{sec:evaluation:stability}

To evaluate whether repeated migration accumulates computational, memory, or continuation debt, we conducted an accelerated long-horizon experiment on the two-node testbed. We triggered a migration every 30 seconds for 300 consecutive handovers, resulting in approximately 2.5 hours of continuous operation. This cadence is deliberately more demanding than the expected orbital handover frequency. Nevertheless, every migration completed before the following trigger.

\begin{table}[h]
\centering
\caption{Operational stability and longevity metrics across recurrent handovers, comparing overall execution with the initial and final handover windows.}
\label{tab:operational_longevity_metrics}
\resizebox{\linewidth}{!}{%
\begin{tabular}{@{}lrrrrr@{}}
\toprule
\textbf{Metric} & \textbf{Median} & \textbf{$p_{95}$} & \textbf{First 50} & \textbf{Last 50} & \textbf{Change} \\
\midrule
Mig. time ($t_{mig}^{\mathcal{M}}$)  & 4,241~ms   & 5,535~ms   & 4,219~ms   & 4,181~ms   & $-38$~ms \\
Downtime                             & 5.52~ms    & 5.96~ms    & 3.76~ms    & 5.51~ms    & $+1.75$~ms \\
CPU (core-s)                         & 0.748      & 0.938      & 0.733      & 0.738      & $+0.005$ \\
Memory                               & 142.5~MB   & 143.8~MB   & 142.2~MB   & 142.9~MB   & $+0.65$~MB \\
Base transfer                        & 135.5~MB   & 135.7~MB   & 135.5~MB   & 135.5~MB   & $-0.01$~MB \\
Event transfer                       & 5.29~MB    & 7.04~MB    & 5.22~MB    & 5.27~MB    & $+0.05$~MB \\
Overhead ($\phi_{mig}$)              & 14.083~\%  & 18.482~\%  & 14.067~\%  & 13.871~\%  & $-0.196$~\% \\
Availability ($A_d$)                 & 99.982~\%  & 99.995~\%  & 99.982~\%  & 99.982~\%  & $+0.000$~\% \\
Duplications                         & 0          & 0          & 0          & 0          & 0 \\
\bottomrule
\end{tabular}%
}
\end{table}

Figure~\ref{fig:evaluation:recurrent_migration} and Table~\ref{tab:operational_longevity_metrics} show no material drift in migration time, CPU consumption, memory, or continuation traffic. Although median downtime increased by 1.75~ms, it remained below 6~ms and did not increase with migration time. Migration time remained stable at a median of 4.24~s ($p_{95}=5.54$~s), while CPU consumption remained at 0.75 core-seconds ($p_{95}=0.94$). Post-cleanup memory had a median of 142.5~MB and increased by only 0.27~MB per 100 handovers, corresponding to less than 0.6~\% over the complete run. Continuation traffic was similarly stable, with a median of 5.29~MB compared with 135.47~MB for the base snapshot, and exhibited no positive trend. Median and $p_{95}$ downtime were 5.52~ms and 5.96~ms, respectively; aggregating all interruptions over the experiment yields $A_d=99.9857\%$. These results indicate that recurrent handovers neither increase migration cost nor accumulate unresolved continuation state, supporting the claim that virtual stationarity can sustain steady-state operation over repeated migrations.

\subsection{Semantic Continuity}
\label{sec:evaluation:continuity}

We evaluate state, effect, and capability continuity across 100 recurrent handovers. The workload incremented a sequence number, obtained a nondeterministic value through \texttt{random\_get}, incorporated both into a rolling 64-bit hash, and emitted the resulting tuple through \texttt{fd\_write}. During Replay, the target reconstructed recorded returns and memory mutations without invoking the output binding; after commit, execution continued through the target-local binding. An offline verifier combined the outputs from both hosts and checked the complete sequence and hash chain.

Across 100 handovers and 176,801 replayed events, the verifier found no sequence gaps, duplicate outputs, or hash-chain mismatches. The gap-free sequence demonstrates state continuity, while the absence of duplicates provides evidence of effect continuity and authority exclusivity for the evaluated output path. Because every hash depends on a nondeterministic \texttt{random\_get} result, the intact chain shows that $\mathcal{P}$ reproduced both system-call outcomes and their memory mutations. Continued output through the target-local \texttt{fd\_write} binding demonstrates capability continuity under $\mathcal{V}$ for the evaluated workload.

\subsection{Scaling and Convergence Bounds}
\label{sec:evaluation:scaling_and_convergence_bounds}

Equation~\ref{eq:migration_time} predicts that Continuous State Projection scales with base-state volume at low event rates and converges only while $r_{sys}<\mu$. We evaluate both properties against checkpoint/restore and pre-copy on two physical nodes under realistic \ac{ISL} profiles, varying base-state volume, memory-write rate, and intercepted host-call rate. Each intercepted call produces one continuation event, so its rate corresponds to $r_{sys}$. Post-copy is omitted because its predecessor-dependent completion times exceed the recurrent handover budgets considered here.

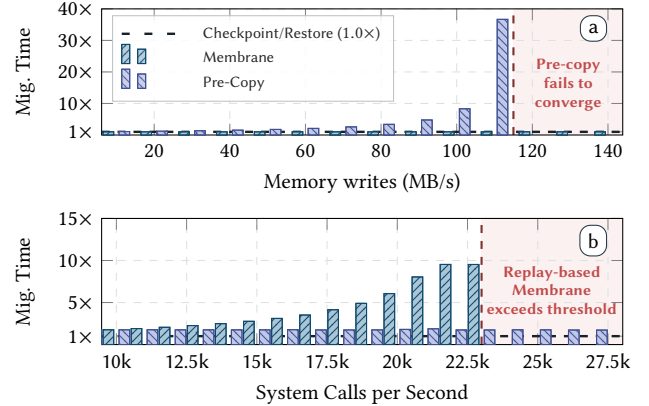
\begin{figure}[h]
    \centering
    \captionsetup[subfigure]{skip=0pt,hypcap=false}

    \begin{subfigure}[t]{\linewidth}
        \phantomcaption
        \label{fig:impact_of_memory_on_migration_duration_impact_memory_write_rate}%
        \begin{tikzpicture}
\begin{axis}[
    ybar,
    width=\linewidth,
    height=3.25cm,
    xlabel={Memory writes (MB/s)},
    ylabel={Mig. Time},
    ymin=0, ymax=40,
    ytick={1, 10, 20, 30, 40, 50},
    yticklabels={$1\times$, $10\times$, $20\times$, $30\times$, $40\times$, $50\times$},
    yticklabel style={
        text width=1.75em,
        align=right
    },
    xlabel style={font=\small},
    ylabel style={font=\small},
    tick label style={font=\sffamily\small},
    xmin=10, xmax=140,
    xtick={20, 40, 60, 80, 100, 120, 140},
    grid=major,
    tick align=inside,
    grid style={dashed, CustomAxes!15},
    bar width=4pt,
    enlarge x limits=0.03,
    legend style={
        at={(0.02, 0.95)},
        anchor=north west,
        font=\sffamily\scriptsize,
        fill=white,
        fill opacity=0.85,
        draw=CustomAxes!30,
        thin,
        row sep=-1pt,
        legend columns=1,
        legend cell align={left},
        /tikz/every odd column/.append style={column sep=8pt},
        /tikz/every even column/.append style={column sep=10pt}
    },
    set layers
]

    \begin{pgfonlayer}{axis background}
        \fill[CustomAlert!8] (axis cs:115,0) rectangle (axis cs:144,40);
        \draw[CustomAlert!60!black, dashed, thick] 
            ({axis cs:115,0} |- {rel axis cs:0,0}) -- ({axis cs:115,0} |- {rel axis cs:0,1});
    \end{pgfonlayer}
    
    \node[
        anchor=center,
        align=center,
        text=CustomAlert!90!black,
        font=\sffamily\scriptsize\bfseries,
    ] at (axis cs:129.5, 16) {Pre-copy\\fails to\\converge};

    \draw[
        CustomAxes!60!black, 
        dashed, 
        line width=0.9pt
    ] ({rel axis cs:0,0} |- {axis cs:10, 1.0}) -- ({rel axis cs:1,0} |- {axis cs:10, 1.0});
    
    \addlegendimage{line legend, CustomAxes!60!black, dashed, line width=0.9pt}
    \addlegendentry{Checkpoint/Restore ($1.0\times$)}

    \addplot[
        fill=CustomPrimary!35,
        draw=CustomPrimary!80!black,
        line width=0.4pt,
        postaction={
            pattern=north east lines,
            pattern color=CustomPrimary!90!black
        }
    ] table [x=rate, y=membrane_factor, col sep=comma] {evaluation/experiments/baseline-compare/normalized_data_rate.csv};
    \addlegendentry{Membrane}

    \addplot[
        fill=CustomChart3!35,
        draw=CustomChart3!80!black,
        line width=0.4pt,
        postaction={
            pattern=north west lines,
            pattern color=CustomChart3!90!black
        }
    ] table [x=rate, y=precopy_factor, col sep=comma] {evaluation/experiments/baseline-compare/normalized_data_rate.csv};
    \addlegendentry{Pre-Copy}

    \node[
        anchor=north east,
        font=\sffamily\small,
        fill=white,
        draw=CustomAxes,
        rounded corners=4pt,
        inner sep=3pt
    ] at (rel axis cs:0.97,0.95) {a};

\end{axis}
\end{tikzpicture} %
    \end{subfigure}

    \begin{subfigure}[t]{\linewidth}
        \phantomcaption
        \label{fig:evaluation:syscall-limit}%
        \begin{tikzpicture}
\begin{axis}[
    ybar,
    width=\linewidth,
    height=3.25cm,
    xlabel={System Calls per Second},
    ylabel={Mig. Time},
    ymin=0, ymax=15,
    ytick={1, 5, 10, 15},
    yticklabels={$1\times$, $5\times$, $10\times$, $15\times$},
    yticklabel style={
        text width=1.75em,
        align=right
    },
    xlabel style={font=\small},
    ylabel style={font=\small},
    tick label style={font=\sffamily\small},
    xmin=10000, xmax=27500,
    xtick={5000, 10000, 12500, 15000, 17500, 20000, 22500, 25000, 27500},
    xticklabels={5k, 10k, 12.5k, 15k, 17.5k, 20k, 22.5k, 25k, 27.5k},
    scaled x ticks=false,
    grid=major,
    tick align=inside,
    grid style={dashed, CustomAxes!15},
    bar width=4pt,
    enlarge x limits=0.03,
    legend style={
        at={(0.02, 0.95)},
        anchor=north west,
        font=\sffamily\scriptsize,
        fill=white,
        fill opacity=0.85,
        draw=CustomAxes!30,
        thin,
        row sep=-1pt,
        legend columns=1,
        legend cell align={left},
        /tikz/every odd column/.append style={column sep=8pt},
        /tikz/every even column/.append style={column sep=10pt}
    },
    set layers,
    legend to name=legendbaseline1,
]

    \begin{pgfonlayer}{axis background}
        \fill[CustomAlert!8] (axis cs:23000,0) rectangle (axis cs:28000,15);
        \draw[CustomAlert!60!black, dashed, thick] 
            ({axis cs:23000,0} |- {rel axis cs:0,0}) -- ({axis cs:23000,0} |- {rel axis cs:0,1});
    \end{pgfonlayer}
    
    \node[
        anchor=center,
        align=center,
        text=CustomAlert!90!black,
        font=\sffamily\scriptsize\bfseries,
    ] at (axis cs:25500, 6.5) {Replay-based\\Membrane\\exceeds threshold};

    \draw[
        CustomAxes!60!black, 
        dashed, 
        line width=0.9pt
    ] ({rel axis cs:0,0} |- {axis cs:10, 1.0}) -- ({rel axis cs:1,0} |- {axis cs:10, 1.0});
    
    \addlegendimage{line legend, CustomAxes!60!black, dashed, line width=0.9pt}
    \addlegendentry{Checkpoint/Restore ($1.0\times$)}

    \addplot[
        fill=CustomPrimary!35,
        draw=CustomPrimary!80!black,
        line width=0.4pt,
        postaction={
            pattern=north east lines,
            pattern color=CustomPrimary!90!black
        }
    ] table [x=rate, y=membrane_factor, col sep=comma] {evaluation/experiments/syscall-limit/normalized_data_rate.csv};
    \addlegendentry{Membrane}

    \addplot[
        fill=CustomChart3!35,
        draw=CustomChart3!80!black,
        line width=0.4pt,
        postaction={
            pattern=north west lines,
            pattern color=CustomChart3!90!black
        }
    ] table [x=rate, y=precopy_factor, col sep=comma] {evaluation/experiments/syscall-limit/normalized_data_rate.csv};
    \addlegendentry{Pre-Copy}

    \node[
        anchor=north east,
        font=\sffamily\small,
        fill=white,
        draw=CustomAxes,
        rounded corners=4pt,
        inner sep=3pt
    ] at (rel axis cs:0.97,0.95) {b};

\end{axis}
\end{tikzpicture}%
    \end{subfigure}

    \caption{Continuous State Projection bounds. (\subref{fig:impact_of_memory_on_migration_duration_impact_memory_write_rate})~Pre-copy fails as memory writes approach transfer capacity, whereas membrane migration remains invariant. (\subref{fig:evaluation:syscall-limit})~Replay-based membrane migration grows as event production approaches replay capacity and eventually exceeds the reference budget.}
    \label{fig:impact_of_memory_on_migration_duration}
\end{figure}

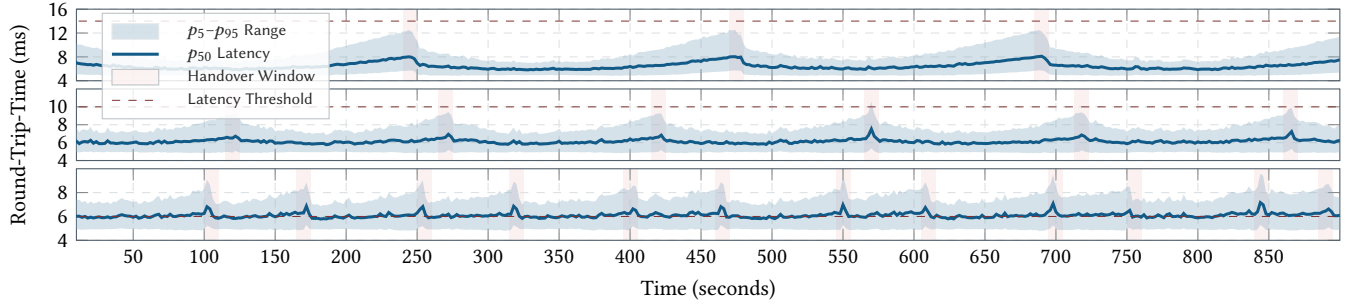
\begin{figure*}[t]
    \centering
    \resizebox{\linewidth}{!}{%
        \begin{tikzpicture}
\begin{groupplot}[
    group style={
        group size=1 by 3,
        vertical sep=3pt
    },
    width=\linewidth,
    height=2.5cm,
    grid=major,
    grid style={dashed, CustomAxes!15},
    xlabel={Time (seconds)},
    xmin=10, xmax=900,
    xtick={50,100,150,200,250,300,350,400,450,500,550,600,650,700,750,800,850},
    ymin=4, ymax=20,
    yticklabel style={/pgf/number format/fixed, /pgf/number format/1000 sep={}},
    xlabel style={font=\small},
    ylabel style={font=\small},
    tick label style={font=\sffamily\small},
    axis line style={CustomAxes!80},
    set layers,
    legend style={
        at={(0.02, 0.95)},
        anchor=north west,
        font=\sffamily\scriptsize,                  
        fill=white,
        fill opacity=0.85,
        draw=CustomAxes!30,
        thin,
        row sep=-1pt,
        legend columns=1,
        legend cell align={left},                    
        /tikz/every odd column/.append style={column sep=8pt},   
        /tikz/every even column/.append style={column sep=10pt} 
    },
]
    \nextgroupplot[
        xlabel={},
        xticklabels={},
        ylabel={},
        ytick={4, 8, 12, 16},
        yticklabels={4, 8, 12, 16},
        ymax=16
    ]

    \begin{pgfonlayer}{axis background}
        \fill[CustomAlert!8] (axis cs:240,4) rectangle (axis cs:250,16);
        \fill[CustomAlert!8] (axis cs:470,4) rectangle (axis cs:480,16);
        \fill[CustomAlert!8] (axis cs:685,4) rectangle (axis cs:695,16);
    \end{pgfonlayer}

    \addplot[name path=upper1, draw=none, forget plot] 
        table [col sep=comma, x=step, y=p95] 
        {evaluation/experiments/macro/paretonormal/14ms_900s_macro_latencies.csv};
    \addplot[name path=lower1, draw=none, forget plot] 
        table [col sep=comma, x=step, y=p5] 
        {evaluation/experiments/macro/paretonormal/14ms_900s_macro_latencies.csv};
    \addplot[fill=CustomPrimary!50, draw=none, opacity=0.35] 
        fill between[of=upper1 and lower1];
    \addlegendentry{$p_5\text{--}p_{95}$ Range}

    \addplot[color=CustomPrimary, line width=1.1pt] 
        table [col sep=comma, x=step, y=median] 
        {evaluation/experiments/macro/paretonormal/14ms_900s_macro_latencies.csv};
    \addlegendentry{$p_{50}$ Latency}

    \addplot[fill=CustomAlert!8, draw=none, area legend] coordinates {(0,0) (0,0)};
    \addlegendentry{Handover Window}

    \addplot[CustomAlert!60!black, very thin, dashed] coordinates {(0,0) (0,0)};
    \addlegendentry{Latency Threshold}

    \draw[CustomAlert!60!black, very thin, dashed] (axis cs:0,14) -- (axis cs:900,14);

    \nextgroupplot[
        xlabel={},
        xticklabels={},
        ylabel={Round-Trip-Time (ms)},
        ytick={4, 6, 8, 10},
        yticklabels={4, 6, 8, 10},
        ymax=12
    ]

    \begin{pgfonlayer}{axis background}
        \fill[CustomAlert!8] (axis cs:115,4) rectangle (axis cs:125,12);
        \fill[CustomAlert!8] (axis cs:265,4) rectangle (axis cs:275,12);
        \fill[CustomAlert!8] (axis cs:415,4) rectangle (axis cs:425,12);
        \fill[CustomAlert!8] (axis cs:565,4) rectangle (axis cs:575,12);
        \fill[CustomAlert!8] (axis cs:713,4) rectangle (axis cs:723,12);
        \fill[CustomAlert!8] (axis cs:860,4) rectangle (axis cs:870,12);
    \end{pgfonlayer}

    \addplot[name path=upper2, draw=none, forget plot] 
        table [col sep=comma, x=step, y=p95] 
        {evaluation/experiments/macro/paretonormal/10ms_900s_macro_latencies.csv};
    \addplot[name path=lower2, draw=none, forget plot] 
        table [col sep=comma, x=step, y=p5] 
        {evaluation/experiments/macro/paretonormal/10ms_900s_macro_latencies.csv};
    \addplot[fill=CustomPrimary!50, draw=none, opacity=0.35] 
        fill between[of=upper2 and lower2];

    \addplot[color=CustomPrimary, line width=1.1pt] 
        table [col sep=comma, x=step, y=median] 
        {evaluation/experiments/macro/paretonormal/10ms_900s_macro_latencies.csv};

    \draw[CustomAlert!60!black, very thin, dashed] (axis cs:0,10) -- (axis cs:900,10);

    \nextgroupplot[
        xlabel={Time (seconds)},
        ytick={4, 6, 8},
        yticklabels={4, 6, 8},
        ymax=10
    ]

    \begin{pgfonlayer}{axis background}
        \fill[CustomAlert!8] (axis cs:100,4) rectangle (axis cs:110,10);
        \fill[CustomAlert!8] (axis cs:165,4) rectangle (axis cs:175,10);
        \fill[CustomAlert!8] (axis cs:250,4) rectangle (axis cs:260,10);
        \fill[CustomAlert!8] (axis cs:315,4) rectangle (axis cs:325,10);
        \fill[CustomAlert!8] (axis cs:395,4) rectangle (axis cs:405,10);
        \fill[CustomAlert!8] (axis cs:460,4) rectangle (axis cs:470,10);
        \fill[CustomAlert!8] (axis cs:545,4) rectangle (axis cs:555,10);
        \fill[CustomAlert!8] (axis cs:605,4) rectangle (axis cs:615,10);
        \fill[CustomAlert!8] (axis cs:695,4) rectangle (axis cs:705,10);
        \fill[CustomAlert!8] (axis cs:750,4) rectangle (axis cs:760,10);
        \fill[CustomAlert!8] (axis cs:840,4) rectangle (axis cs:850,10);
        \fill[CustomAlert!8] (axis cs:885,4) rectangle (axis cs:895,10);
    \end{pgfonlayer}

    \addplot[name path=upper3, draw=none, forget plot] 
        table [col sep=comma, x=step, y=p95] 
        {evaluation/experiments/macro/paretonormal/6ms_900s_macro_latencies.csv};
    \addplot[name path=lower3, draw=none, forget plot] 
        table [col sep=comma, x=step, y=p5] 
        {evaluation/experiments/macro/paretonormal/6ms_900s_macro_latencies.csv};
    \addplot[fill=CustomPrimary!50, draw=none, opacity=0.35] 
        fill between[of=upper3 and lower3];

    \addplot[color=CustomPrimary, line width=1.1pt] 
        table [col sep=comma, x=step, y=median] 
        {evaluation/experiments/macro/paretonormal/6ms_900s_macro_latencies.csv};

    \draw[CustomAlert!60!black, very thin, dashed] (axis cs:0,6) -- (axis cs:900,6);

\end{groupplot}
\end{tikzpicture}%
    }
    
    \caption{End-to-end latency during continuous migration for 14~ms, 10~ms, and 6~ms target \acp{SLO} on a 1,600-satellite constellation. Blue lines show median \ac{RTT}; tighter targets shorten the median interval between handovers to 242~s, 151~s, and 83~s, respectively.}
    \label{fig:evaluation:membrane_latency_macro_grouped}
\end{figure*}

Figure~\ref{fig:impact_of_memory_on_migration_duration_impact_memory_write_rate} shows that membrane migration varies by less than 1\% across write loads of 10--140~MB/s, whereas pre-copy ceases to converge as the dirtying rate approaches available bandwidth. In a separate 200--1{,}000~MB base-state sweep, membrane and checkpoint/restore yield linear slopes of 9.06 and 8.96~ms/MB, respectively ($R^2>0.999$), corresponding to an effective bandwidth of $C\approx110$~MB/s. Across this sweep, membrane downtime remains within 2--5~ms. On the other hand, Figure~\ref{fig:evaluation:syscall-limit} confirms the event-rate bound predicted by Equation~\ref{eq:migration_time}. Fixing $\Phi/C$ from the base-state experiment and fitting the remaining parameters yields $\mu=24.1\times10^{3}$~events/s and $t_{ovh}=136$~ms. Measured durations deviate from the model by 1.3\% on average over 1{,}000--21{,}000~calls/s, while a fit trained on 1k--15k~calls/s predicts the withheld 16k--21k measurements within 0.8\% mean error. Using the 15~s beam-switch period reported by Zhao et al.~\cite{zhao_demystifying_2026} as a demanding reference rather than a universal budget $B_k$, the model predicts a breach at 20.2k~calls/s. Here, $\mu$ denotes the end-to-end catch-up rate, including event ingestion, application of $\Delta M$, and deterministic execution between events.

\subsection{Portability}
\label{sec:evaluation:portability}

The framework requires its Execution Engine to provide portable execution ($\mathcal{X}$). We validate this requirement through bidirectional migration between an x86-64 host (Lenovo ThinkPad, 4.31~GHz, 32~GB RAM) and an ARM64 node (Raspberry Pi~5). In all $N=100$ trials, the workload resumed on the target without application changes. The result confirms that the selected WAMR engine satisfies $\mathcal{X}$ for the evaluated architecture pair; it does not assume that arbitrary native runtimes are architecture-independent.

\subsection{Orbital Feasibility}
\label{sec:evaluation:orbital_feasibility}

We evaluate repeated handovers over multi-minute runs using a constellation simulator based on the \ac{SGP4} model~\cite{vallado_sgp4_2008}. The simulator calculates satellite trajectories and supplies the migration schedule and corresponding \ac{ISL}/\ac{GSL} profiles to the physical testbed. Two Raspberry Pi 5 nodes alternate as source and successor. This experiment evaluates the framework’s runtime lifecycle under an externally supplied schedule, rather than the optimality of successor selection.

\begin{table}[h]
\centering
\caption{Orbital runtime metrics (medians); $A_d$ includes runtime interruption only.}
\label{tab:migration_tax_summary}
\small
\resizebox{\linewidth}{!}{%
\begin{tabular}{@{}l r r r r r@{}}
\toprule
\textbf{Latency} &
\textbf{Mig. Time} &
\textbf{Downtime} &
\textbf{On Host} &
\textbf{Overh. ($\phi_{mig}$)} &
\textbf{Avail. ($A_d$)} \\
\midrule
14~ms & 3212~ms & 4.09~ms & 242~s & 1.339~\% & 99.998~\% \\
10~ms & 3229~ms & 3.85~ms & 151~s & 2.267~\% & 99.997~\% \\
 6~ms & 3402~ms & 2.10~ms &  83~s & 4.403~\% & 99.997~\% \\
\bottomrule
\end{tabular}%
}
\end{table}

We evaluate target latency profiles of 14, 10, and 6~ms, with the latter approaching the physical lower bound of the modeled links. Figure~\ref{fig:evaluation:membrane_latency_macro_grouped} shows an active \ac{XR} workload receiving approximately 4{,}000 packets/s. Across the 900~s traces, median \ac{RTT} exhibits no upward drift over successive handovers. Migrations complete in 2.8--4.5~s within observed handover intervals of 83--242~s, leaving substantial budget under the condition $t_{mig,k}\leq B_k$. The similar migration times across profiles show that tighter latency targets increase handover frequency rather than individual migration cost.

The same logical UDP descriptor is rebound and remains usable across every handover, demonstrating capability continuity. Across 20 aligned handovers, the time-averaged $p_{50}$ packet-loss trace is 2.01\% during the handover window, compared with 1.97\% before and 1.99\% after. Its maximum within-window $p_{50}$ is 2.22\%, while $p_{95}$ peaks at 5.10\%. Median interruptions of 2.10--4.09~ms yield $A_d\geq99.997\%$, while $\phi_{mig}$ ranges from 1.339\% to 4.403\% (Table~\ref{tab:migration_tax_summary}). Because $A_d$ excludes external path updates and ordinary latency violations, it does not establish end-to-end availability $A$.

\section{Conclusion \& Future Work}
\label{sec:conclusion}

We presented \ac{VSR}, a runtime framework that treats recurrent orbital migration as a steady-state execution primitive. Starting from a mechanism-independent feasibility model, the framework separates external host selection and scheduling from runtime handover. Its Migration Membrane combines Capability Virtualization to rebind host-dependent resources with Continuous State Projection to reconstruct an evolving execution on the successor. An explicit authority fence completes the handover without retaining predecessor dependence or permitting both hosts to produce external effects. Our prototype demonstrated that this model can operate across heterogeneous hosts. Across 300 recurrent handovers, $p_{95}$ downtime remained below 6~ms without cumulative growth in migration time or continuation state. The prototype also preserved the evaluated state, capability, and effect continuity properties and resumed successfully across x86-64 and ARM64. Under simulated orbital schedules, migrations completed within the available handover intervals while maintaining bounded migration occupancy and downtime-only availability of $A_d\ge99.997\%$.

Future work will incorporate orbital, network, energy, and thermal constraints into external host selection. We also plan to investigate continuously active state projection for constellations requiring higher handover frequencies and broader classes of virtualized capabilities.


%
%

%
%


\bibliographystyle{ACM-Reference-Format}
\bibliography{references}

\end{document}